\documentclass[iop]{emulateapj}

\usepackage{aas_macros}
\usepackage{multirow}

\newcommand{\satmc}{M_{\rm{c}}^r}
\newcommand{\satdmbin}{\Delta M_{\rm{bin}}}
\newcommand{\satdmf}{\Delta M_{\rm{faint}}}
\newcommand{\satdzs}{\Delta z_{\rm{s}}}
\newcommand{\satap}{\alpha_{\rm{p}}}

\newcommand{\ri}{R_{\rm{inner}}}
\newcommand{\ro}{R_{\rm{outer}}}

\newcommand{\lcdm}{$\rm{\Lambda CDM}$~}

\newcommand\lsim{\mathrel{\hbox{\rlap{\hbox{\lower4pt\hbox{$\sim$}}}\hbox{$<$}}}}

\shorttitle{Satellite LFs of Galaxies in Filaments}
\shortauthors{Guo et al.}

\begin{document}

\title{Galaxies in filaments have more satellites: the influence of the cosmic web on the satellite luminosity function in the SDSS}

\author{Quan Guo}
\affil{Leibniz-Institut f\"ur Astrophysik Potsdam, An der Sternwarte 16,
D-14482 Potsdam, Germany}
\email{qguo@aip.de}
\author{E. Tempel}
\affil{Tartu Observatory, Observatooriumi~1, 61602 T\~oravere, Estonia}
\affil{National Institute of Chemical Physics and Biophysics, R\"avala pst 10, Tallinn 10143, Estonia}
\author{N. I. Libeskind}
\affil{Leibniz-Institut f\"ur Astrophysik Potsdam, An der Sternwarte 16,
D-14482 Potsdam, Germany}

\label{firstpage}

\begin{abstract}
  We investigate whether the satellite luminosity function (LF) of primary
  galaxies identified in the Sloan Digital Sky Survey (SDSS) depends on whether
  the host galaxy is in a filament or not. Isolated primary galaxies are
  identified in the SDSS spectroscopic sample while potential satellites (that
  are up to 4 magnitudes fainter than their hosts) are searched for in the much
  deeper photometric sample. Filaments are constructed from the galaxy
  distribution by the ``Bisous'' process. Isolated primary galaxies are divided
  into two subsamples: those in filaments and those not in filaments. We examine
  the stacked mean satellite LF of both the filament and non-filament sample and
  find that, on average, the satellite LFs of galaxies in filaments is
  significantly higher than those of galaxies not in filaments. The filamentary
  environment can increases the abundance of the brightest satellites
  ($M_\mathrm{sat.} < M_\mathrm{prim.} + 2.0$), by a factor of $\sim 2$ compared
  with non-filament isolated galaxies. This result is independent of primary
  galaxy magnitude although the satellite LF of galaxies in the faintest
  magnitude bin, is too noisy to determine if such a dependence exists. Since
  our filaments are extracted from a spectroscopic flux-limited sample, we
  consider the possibility that the difference in satellite LF is due to a
  redshift, colour or environmental bias, finding these to be insufficient to
  explain our result. The dependence of the satellite LF on the cosmic web
  suggests that the filamentary environment may have a strong effect on the
  efficiency of galaxy formation.
\end{abstract}


\section{Introduction}
  The \lcdm~model predicts that structure forms in a hierarchical manner.  The
  first objects to collapse and virialize at high redshift are small dark matter
  haloes that later merge to form larger objects. Small haloes that host
  satellite or dwarf galaxies, can often survive the violent process associated
  with halo mergers for many Giga-years providing important information about
  galaxy formation, the population of subhaloes, and even the nature of dark
  matter.  
  
  Moreover these dark matter structures form an intricate pattern on the
  megaparsec scale, known as ``cosmic web'' \citep{bon96}, consisting of regions
  termed voids, filaments, sheets and knots. The cosmic web is a direct
  consequence of the gravitational instabilities that emerge out of the
  primordial density field. The presence of such cosmic pattern has been
  confirmed observationally by the distribution of the galaxies from the large
  surveys such as the 2dF Galaxy Redshift Survey \citep[2dFGRS;][]{col01,
  col03},  the Sloan Digital Sky Survey \citep[SDSS;][]{yok00,teg04}, the Two
  Micron All Sky Survey \citep[2MASS;][]{huc05}, Galaxy And Mass Assembly
  \citep[GAMA;][]{alp04} and the CosmicFlow-2 survey of peculiar velocities \citep{Tully:14}. 

A number of studies have examined how the cosmic web can affect specific halo
properties such as abundance, shape, or assembly history
\citep{ara07,hah07a,hah07b,lib12,lib13b,cau13}. Correlations have also been
found between halo spin and the principle axis of filaments (their spine) and
walls (their normal) that they are embedded in \citep{alt06, ara07,
hah07a, hah07b, zha09, lib13, ara13, dub14}.
  Although more difficult to quantify owing to degeneracies and inherent biases,
  similar studies have been conducted in observational samples
  \citep{jon10,tem13a,tem13b,zha13}. At $z=3.1$ \cite{mats04} found that
  the spatial distribution of galaxies within Ly$\alpha$ haloes trace the underlying
  large-scale filamentary structure  of the universe. The
  morphology and spatial extent of Ly$\alpha$ haloes depend on the environment as
  well \citep{mats11, mats12}.
  
  Tying galaxy or halo properties to the environment is not a new idea and many
  studies have quantified the importance of such relations for galaxy formation
  and evolution \citep[e.g.][]{dre80, kau04, bla05b, Tempel:11}. On the larger
  scales of the cosmic web, the properties of galaxies depend on the cosmic
  filament it inhabits \citep[e.g.][]{mur11,jon10} or on the supercluster
  environment \citep[e.g.][]{Lietzen:12,Einasto:14}. However, to date no
  observational studies have examined the effect of the cosmic web environment
  on satellite galaxies.
  
   Analyzing satellite systems of external
  galaxies is challenging, because typically only several satellites are
  detected per primary galaxy. Furthermore the real space position of a satellite with
  respect to its primary is uncertain. Owing to the advent of large
  galaxy surveys, a statistically robust estimate of the satellite luminosity function (LF) has become
  possible \citep[e.g.][]{my11,tal11,wang12}.

  In this study, we investigate if the satellite LF depends on whether the
  host galaxy is located within a filament.  We divide
  isolated primary galaxies into two categories: those within galaxy filaments
  defined by the ``Bisous process'' as in \cite{tem14} and those not in these
  filaments.  Each subsample is then divided into three bins according to the
  primary's magnitude. The resulting subsamples enable us to study the satellite
  LFs of isolated primary galaxies in filaments and compare it with isolated
  primary galaxies  that are not in filaments. Such an analysis will quantify
  how the filament environment affects galaxy formation.
 
  Throughout the paper we assume a fiducial ${\rm\Lambda CDM}$
  cosmological model with $\Omega_{\rm M}=0.3$, $\Omega_{\Lambda}=0.7$ and
  $H_0 = 70\ $km~s$^{-1}$Mpc$^{-1}$.

\section{Data and Methods} Throughout this paper we refer to isolated {\it
  primary} galaxies (or just {\it ``primaries''}) as the central galaxies that
  host systems of fainter satellites and fulfill a set isolation criteria,
  detailed below.

  \subsection{Galaxies and their satellites}
  \label{sec:iso} The first step in our analysis is to identify isolated
  primaries. In order to do so we use the isolated primary catalogue selected by
  \cite{my11, my12}. 

Here, potential primaries are drawn from the SDSS DR8 spectroscopic survey. All
galaxies within a projected distance of $2R_{\rm inner}$ of a potential primary
are examined (see below for the definition of $R_{\rm inner}$). These must be
more than half a magnitude fainter than a prospective host, unless the
spectroscopic redshift difference is greater than 0.002. If a galaxy within
$2R_{\rm inner}$ only has a photometric redshift, than the host is considered
isolated only if the the difference between the host's spectroscopic redshift
and the interloper's photometric redshift is greater than 2.5 times the
photometric error. As a sanity check our isolated sample is cross matched with
the group catalogue of Yang et al (2007) - an insignificant fraction of our
primaries are considered members of groups according to Yang et al (2007).
Omitting these has no discernible effect on our results.  We use de-reddened
$ugriz$ bands model magnitudes and $k$-correct all galaxies to $z=0$ with the
IDL code of \citet{bla07}.

\begin{figure*}
  \includegraphics[width=55mm]{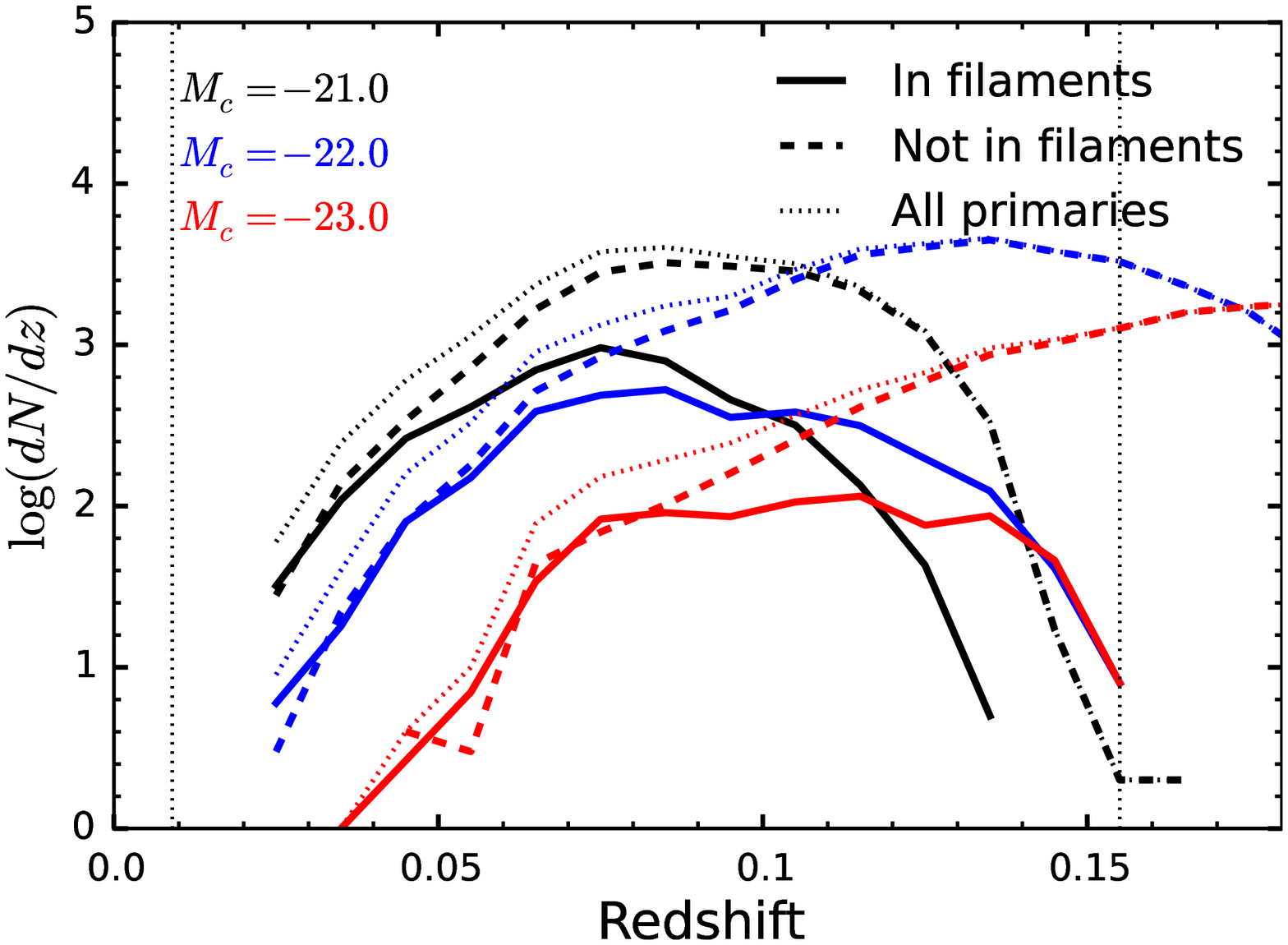}
  \includegraphics[width=55mm]{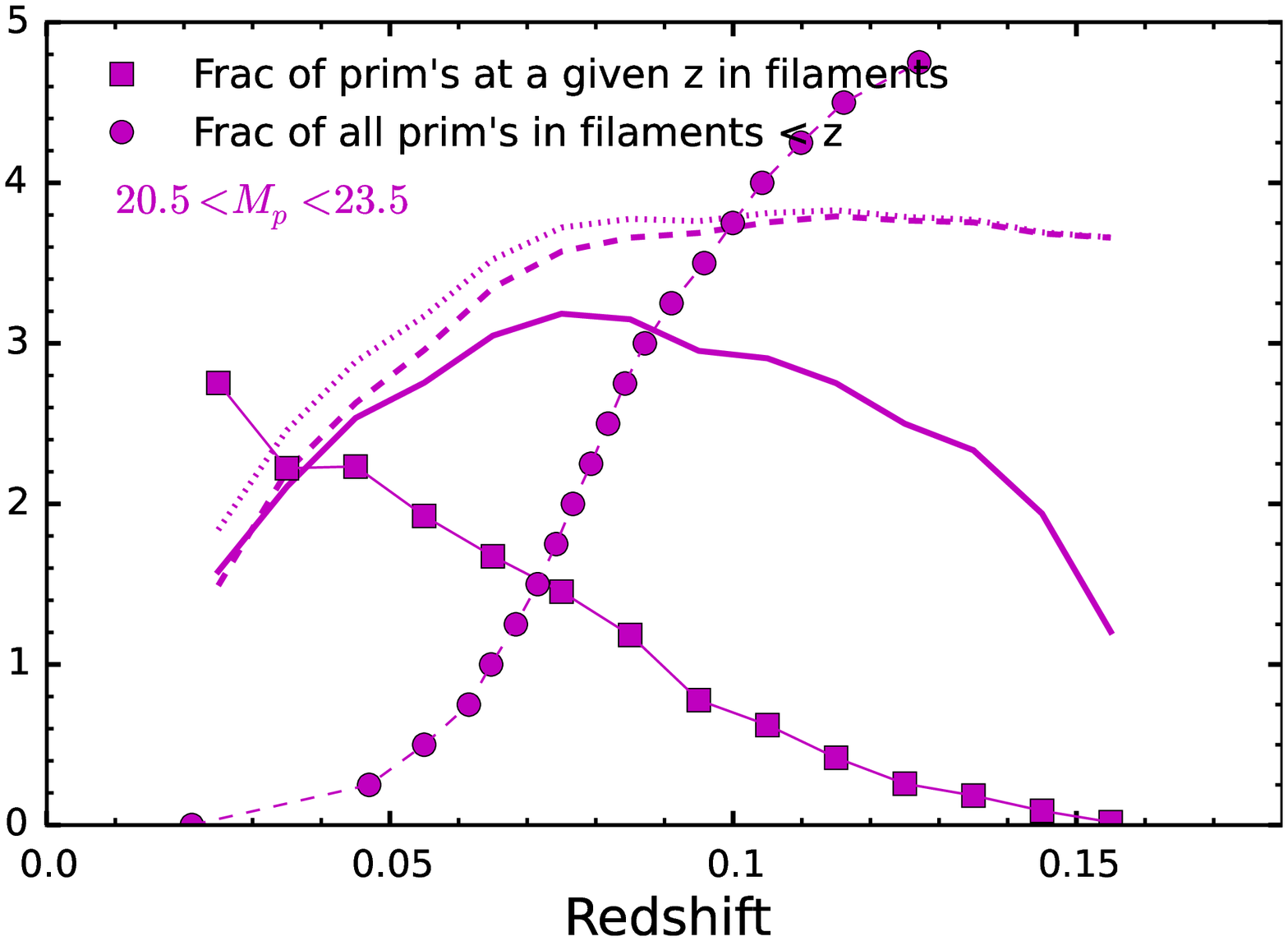}
  \includegraphics[width=55mm]{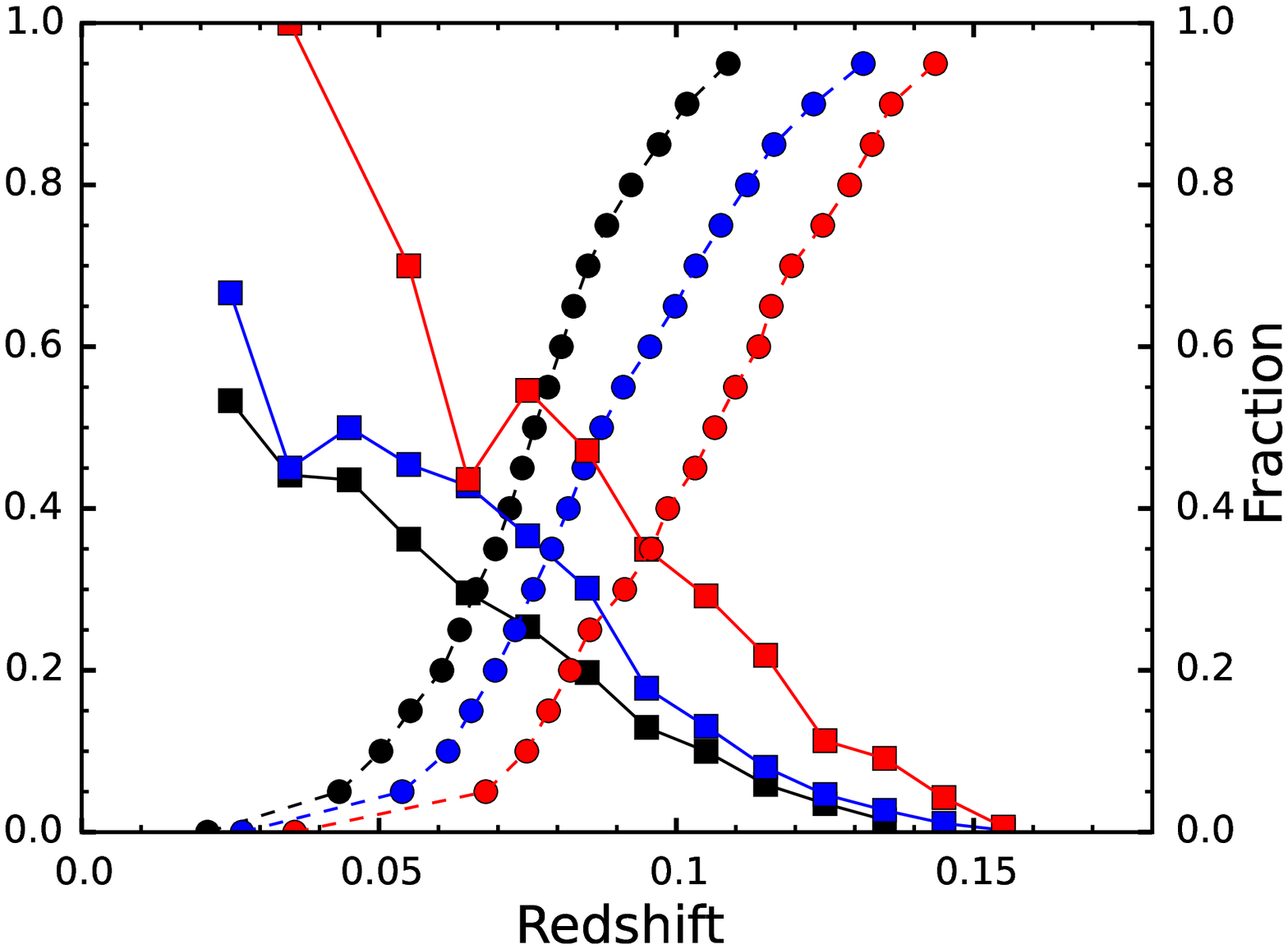}
  \caption{\small {\it Left}: The redshift distribution of our primary sample,
  divided into three magnitude bins centered on $M_{r}=-21.0$, $-22.0$, $-23.0$
  (shown in black, blue, and red respectively). The dotted, dashed and solid
  lines show these distributions for all primaries, those found in filaments and
  those not in filaments (respectively). The two vertical thin dotted lines
  indicate the upper and lower $z$ limit of the filament catalogue. {\it
  Center:} the redshift distribution for all primaries at all magnitudes (i.e.
  $-20.5<M_{r}<-23.5$) is line coded by filamentary environment according to the
  same scheme. The lines connected by squares and circles indicate the fraction
  of primaries at given redshift found in filaments and the and the fraction of
  all primaries that are in filaments below a given $z$. {\it Right:} the same
  two fractions shown divided by magnitude.} \label{fig:fig1_rs}
\end{figure*}

Satellite galaxies are drawn from both spectroscopic and photometric samples
\citep{aih11}. Since the photometric redshifts constrain the distance of a
satellite poorly, it is impossible to properly de-project a potential satellite.
Thus a statistical background subtraction technique is used to estimate the real
satellite galaxy population around each primary (see section~3 of \citet{my11}).
We begin by defining two radii, $\ri$ and $\ro$. $\ri$ represents the projected
radius within which satellites may reside. The annulus $\ro-\ri$ defines the
region within which the local background is estimated. The luminosity function
in this annulus is computed for each primary and then subtracted from the
luminosity function within $\ri$.

The values of $\ri$ and $\ro$ depend on the primary's magnitude. We divide the
primary sample into three $r$-band magnitude bins, each one magnitude wide and
centred on $M_{r}=$ $-21.0$, $-22.0$, and $-23.0$. The values of $(\ri, \ro)$
are (0.3, 0.6), (0.4, 0.8) and (0.55, 0.9)~Mpc, respectively (see \citet{my12}
for details on the choice of these values.) The magnitude bins correspond to
haloes with mean virial radii of around 240, 370, 520~kpc respectively
\citep{my12}.  Finally, the background-subtracted satellite LFs for each
isolated primary in a given absolute magnitude bin is averaged, resulting in an
estimate of the mean satellite LF. \cite{my13} verified that the methodology
described above returns the true satellite LF when applied to galaxy samples in
simulations. 

 We also examine the projected number density profiles of satellites more
 luminous than a particular absolute magnitude \citep[in
 Appendix~\ref{appendix:profile}, as described more fully
 in][]{my11, my12}. Such density profiles are shown in units of $r_{200}$ and
 divided by the total number of satellites within this radius. The values of
 $r_{200}$ are the same as in \cite{my12,my13} and equal to 0.24, 0.37, 0.52~Mpc
 for three magnitude bins respectively.

Finally, the effects of incompleteness and other biases are examined  in
Appendix~\ref{appendix:test}.

  \subsection{SDSS galaxy filaments} \label{sec:fila} The catalogue of filaments
  is built by applying an object point process with interactions (the Bisous
  process) to the distribution of galaxies in the spectroscopic galaxy sample
  from SDSS DR8 as described in \cite{tem14}. Random small segments (cylinders)
  based on the positions of galaxies are used to construct a filamentary network
  by examining the connectivity and alignment of these segments.  A filamentary
  spine can then be extracted based on a detection probability and filament
  orientation \citep[see][]{tem14}.

  The catalogue of filaments (``filament spines'') we use in this study is the
  same that of \cite{tem14}, where the assumed filament radius is roughly
  $0.71~{\rm Mpc}$. This means that primary galaxies in filaments are less than
  0.71~Mpc from the filament spine.  This catalogue is constructed from the full
  spectroscopic galaxy sample with lower and upper CMB-corrected distance limits
  of $z=0.009$ and $z=0.155$ respectively. We thus confine our analysis to this
  redshift range. Filaments at higher redshift than the upper limit are too
  ``diluted'' to be detected. We classify a galaxy as ``in-a-filaments'' if the
  distance of the galaxy from the axis of the filament is less than $0.71~{\rm
  Mpc}$ and the distance of the galaxy from the end point of filament (if the
  galaxy is outside a filamentary cylinder) is less than 0.14~{\rm Mpc}.

\subsection{Redshift biases}

  For each magnitude bin $M_{r}=-21$, $-22$, $-23$ there are 4425 (22795), 3077
  (27857), 740 (4875) primaries in filaments (not in filaments). Thus the
  fraction of isolated galaxies in filaments is roughly 18.5\%, 11.0\% and
  15.1\%, respectively. The (normalized) redshift distribution of primaries (in
  and not in filaments) is shown in the left panel of Figure~\ref{fig:fig1_rs}.
  At a given magnitude, the redshift distribution of galaxies in filaments tends
  to peak at lower $z$ than those not in filaments.  In the right panel of
  Figure~\ref{fig:fig1_rs}  we show the fraction of isolated galaxies in
  filaments as a function of redshift for our three magnitude bins. These vary
  from $\sim$50\% at low $z$ to $\sim0\%$ at the limit of our redshift range.  
  
  Such a dramatic drop is mainly driven by the filament finding algorithm: the
  probability of detecting filaments decreases significantly in the redshift
  range considered here \citep[namely from $z=0.009$ to 0.15, see][]{tem14}.
  This is because filaments are identified in the flux-limited galaxy sample: as
  one goes to higher $z$ the number density of galaxies decreases significantly,
  making the identification of filaments more difficult. Also note that the
  catalogue from which primaries are drawn is also flux-limited: the  number of
  galaxies also decreases with redshift. Thus the combined effect of lower
  number density of isolated galaxies with fewer identifiable filaments at high
  redshift results in the number of isolated galaxies in filaments going to null
  at $z>0.15$. Therefore the dramatic drop of primaries found in filaments at
  higher $z$ is \textit{not} a real feature of the galaxy distribution.
  
Therefore, we exclude galaxies at high ($z>0.15$) and at low ($z<0.04$)
redshifts, since these redshifts are not well covered by the filament catalogue.
We only consider primaries in the redshift ranges $0.04<z<0.13$, $0.04<z<0.14$
and $0.06<z<0.15$ for magnitudes of $-21$, $-22$ and $-23$ respectively, in
order to ensure that the fraction of primaries in filaments is kept at around
($\sim 50\%$).

\section{Results}

\begin{figure} 
  \plotone{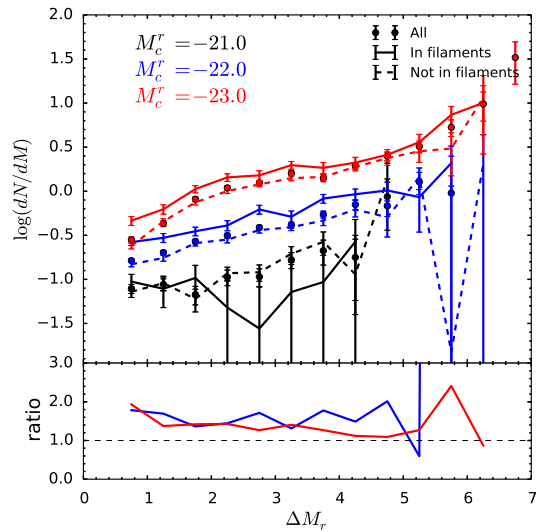} 
  \caption{The satellite LF (top panel) for all primary galaxies,
  those in filaments and those not in filaments is shown as points, solid lines
  and dashed lines, respectively for three different magnitude bins $M_{r}=-21.0$, $-22.0$,
  $-23.0$ (black, blue and red lines, respectively). The ratio of the
  satellite LFs of galaxies in filaments to those in the same magnitude bin but
  not in filaments  is shown in bottom panel (for $M_{r}=-22.0$, $-23.0$).}
  \label{fig:fig2_lf} 
\end{figure}

  The satellite LF for primaries in and not in filaments is shown in top panel
  of Figure~\ref{fig:fig2_lf}. For the magnitude bin $\satmc = -21.0$, the small
  number of primaries results in a LF that is too noisy to show any difference
  (if one existed): the mean number of satellites around these low luminosity
  primary galaxies is intrinsically lower than that of brighter primaries and
  the number of these low-luminosity primaries found in filaments is not
  sufficiently high to have good signal-to-noise ratio.

  Beyond the faintest magnitude bin, the satellites LF of primaries in filaments
  (extending to at least 4 magnitudes fainter than the primary) is significantly
  higher than  those of primaries that are not in filaments. At the bright end
  of the satellite LFs we find that there are roughly double the number of
  satellites around primaries that are in filaments compared  those that are not
  in the filaments. This is our main result: {\it at a given magnitude, isolated
  galaxies in filaments have more bright satellites than isolated galaxies that
  are not in filaments}.

  Since the fractions of galaxies in filaments varies significantly with the
  redshift, we wish to test whether the  difference in the satellite LFs we have
  seen is caused by the bias in the redshift distribution. We thus perform a
  number of tests described in Section~\ref{sect:weighting} to
  \ref{sect:controlsample}. 

  \subsection{Applying weights to the Satellite LFs} 
  \label{sect:weighting}
  Suppose our main result - that galaxies in filaments have {\it on average} more
  satellites -  is driven by the a redshift evolution of the satellite LF combined
  with the decreasing likelihood in finding isolated primaries in filaments. In
  order to control for this possibility, the contribution to the mean
  ``not-in-filament'' satellite LF from a given primary is weighted according to
  what fraction of all galaxies, at that $z$, are found in filaments and what
  fraction are not. For example, at $0.07<z<0.08$ most galaxies (68\%) (in
  magnitude bin $\satmc = -22.0$) are not in filaments. Thus when we compute the
  contribution for a primary that is not found in filaments and in this redshift
  range to the mean LF, we weigh its satellite LF by 0.47 (=32/68).  In this way
  redshift ranges where filaments are easily detected  and redshift ranges where
  filaments are not easily detected are given inverse weights. If the result we
  have found is driven by a redshift dependent satellite LF, such a test should
  give identical satellite LFs for in filament and not-in-filament samples.  The
  satellite LF with weighting is thus estimated as: 
  \begin{equation} 
    \overline{N}_{j}= \frac{\sum\limits_{i=1}^{N_j}W_iN_i(M_j)}{\sum\limits_{i=1}^{N_j}W_i}, 
  \end{equation}
  where $N_i$ is the number of satellites around primary galaxy $i$, $W_i$ is
  the weight for the primary galaxy $i$ and $N_j$ is the number of primaries
  contributing to the $j$th bin of the LF.  The weight $W_i$ for each primary
  galaxy is determined by the aforementioned way.

\begin{figure}
  \plotone{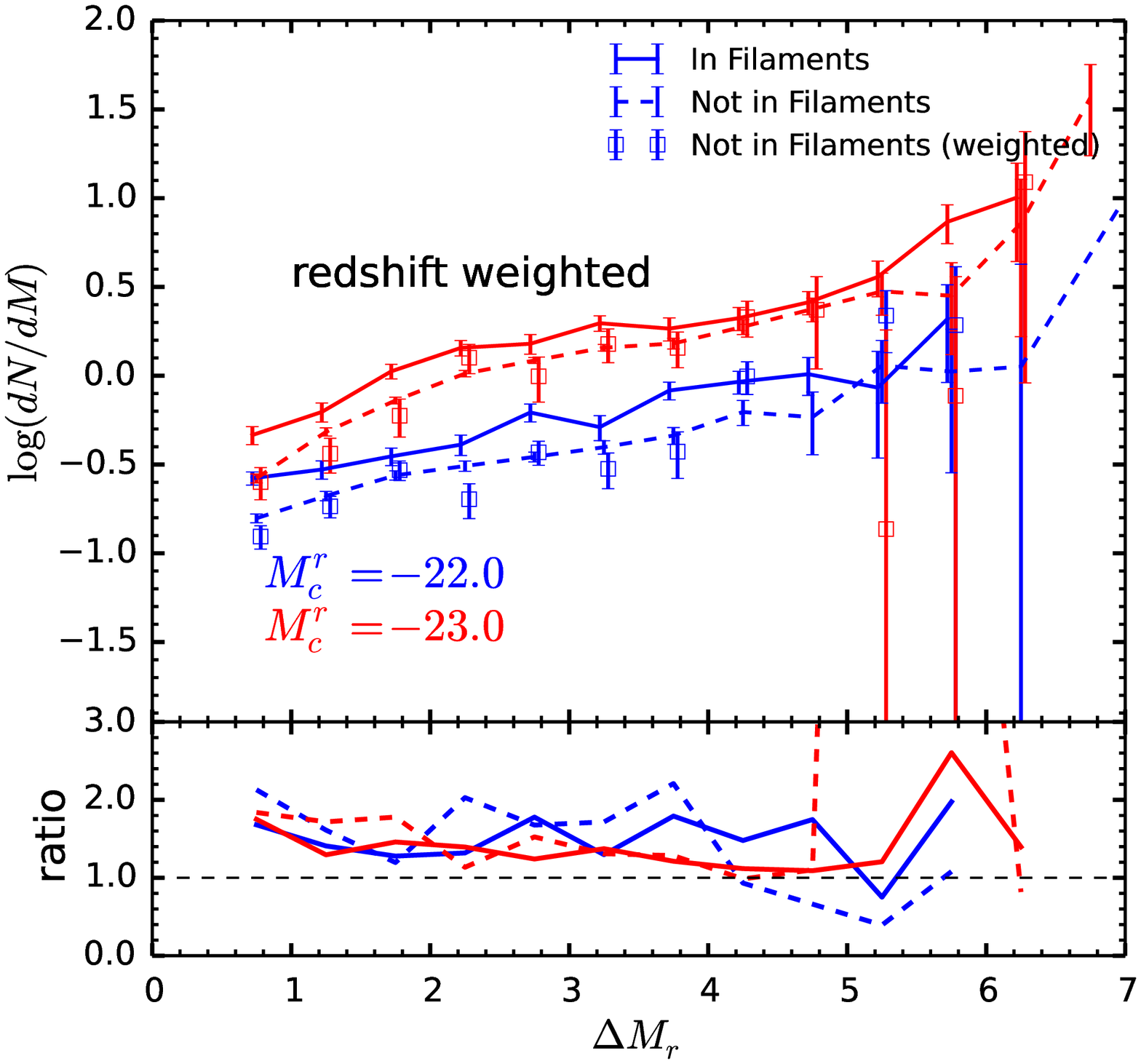}
  \plotone{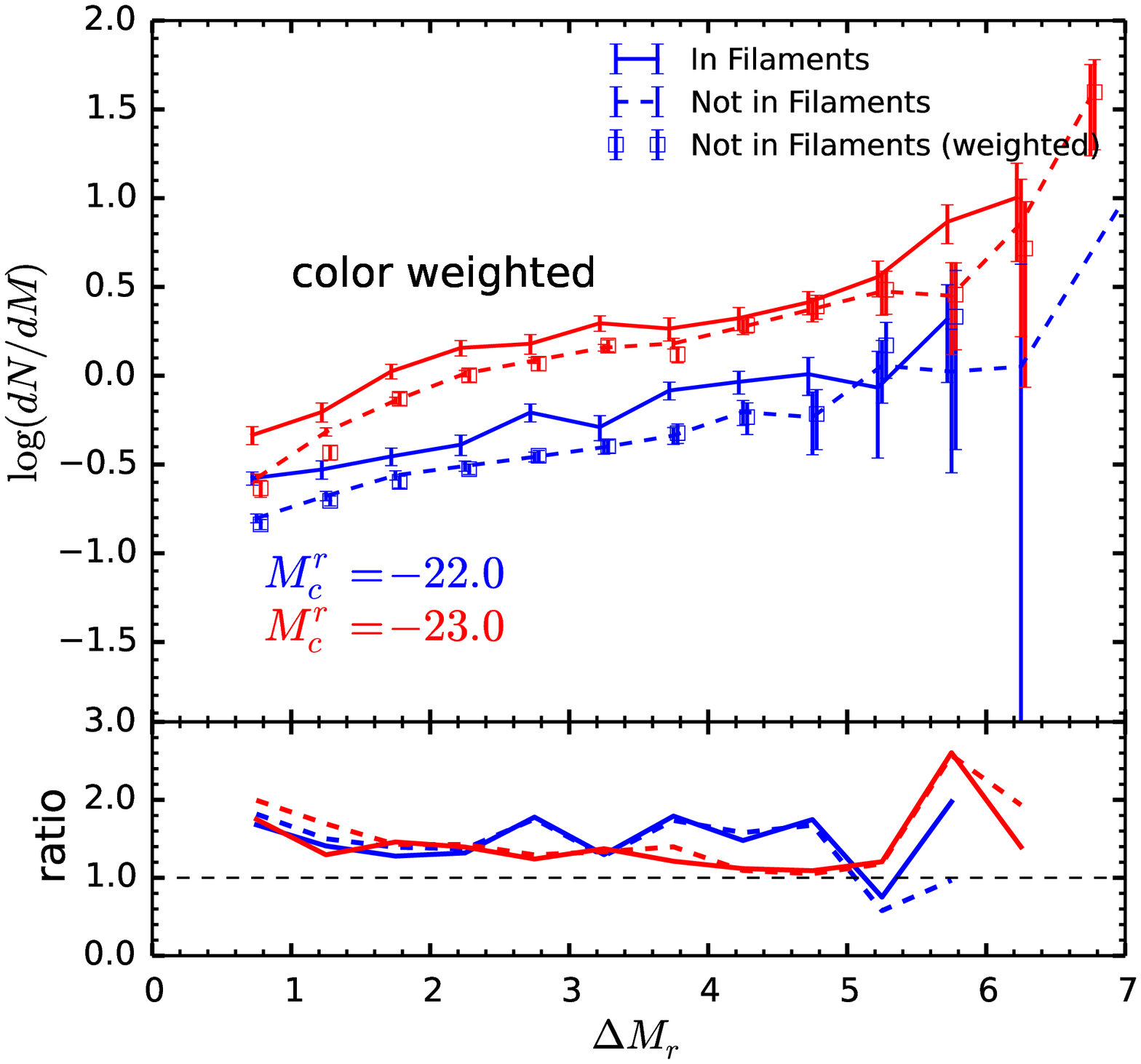}
    \caption{Tests of whether the difference in the mean satellite LF is due
    to the different redshift (top) or color (bottom) distributions of
    filament/not-in-filament primaries, are shown here. Accordingly, the
    contribution to the mean satellite LF due to a not-in-filament primary is
    weighted by the fraction of all primaries (at that $z$, top or with that
    color, bottom) that are found in filaments. The weighted not-in-filament
    satellite LF is shown by squares while the not-in-filament one is shown by
    the dashed lines. The unweighted satellite LF for filament primaries is
    shown for reference (solid lines). The small bottom panels show the ratio
    of the satellite LFs of in-filaments galaxies to those of not-in-filaments
    galaxies as well as the ratio of satellite LFs of in-filaments galaxies to
    the weighted satellite LFs of not-in-filaments galaxies as dashed lines.
    Color coding is the same as in Fig.~\ref{fig:fig2_lf}.}
    \label{fig:fig3_lfw}
\end{figure}

  The top panels of Figure~\ref{fig:fig3_lfw} shows the resulting satellite LFs
  with the aforementioned weighting from the redshift distribution of primary
  galaxies not in filaments for the magnitude bins $\satmc = -22.0$ and $-23.0$.
  The weighted satellites LF for in filament primaries is still significantly
  different from those not in filaments (in a given magnitude bin). Note that
  the weighted satellite LFs for primaries not in filaments are noisier because
  fewer primaries are able to contribute to the estimation of the mean satellite
  LFs.  This suggests that the redshift bias cannot account for the differences
  between the satellite LFs in filaments. 

  Now, suppose our main result is driven by the primary colour distribution. A similarly
  weighted satellite LFs can be estimated in the same way but with weightings
  determined such that the distribution of $(g-r)$ colours (rather than the
  distribution of redshifts) between filament and non-filament primaries match.
  Again, such a LF suggests that a potential colour bias also fails to account for
  the difference in satellite LF (see the bottom panel of
  Figure~\ref{fig:fig3_lfw}).

  Another source of possible systematic bias is the
  different environmental densities around primary galaxies in filaments and not
  in filaments. This has been checked and statistically the primary galaxies in
  filaments and not in filaments have the same distribution of environmental
  densities. In summary: {\it the difference in filament and non-filament
  satellite LF cannot be attributed to an underlying colour, redshift or
  environmental density bias. The filament and non-filament satellite LFs are
  intrinsically different}.

\subsection{Dependence on the redshift of primary galaxies} \label{sec:redshift}

\begin{figure}
    \plotone{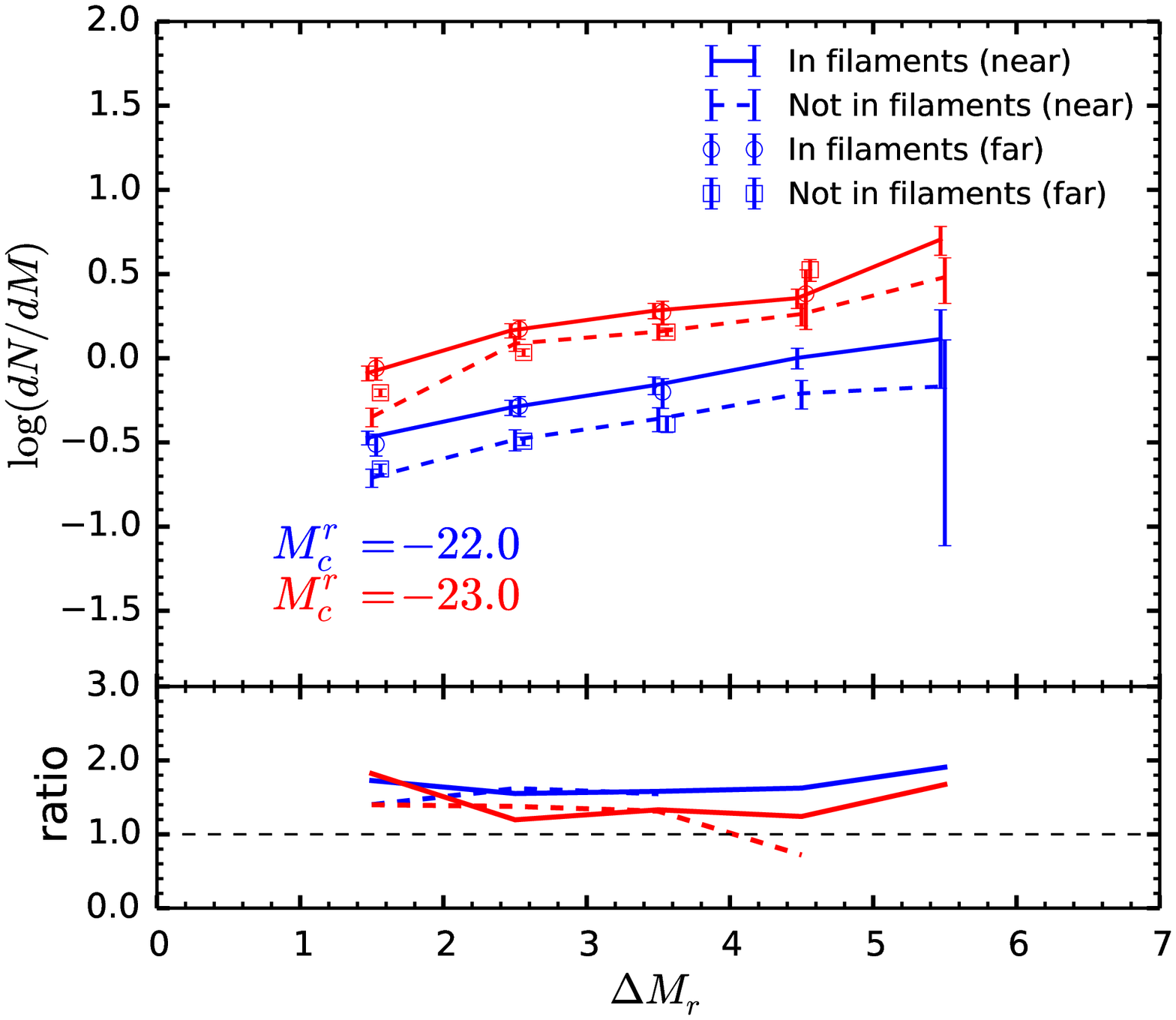}
    \plotone{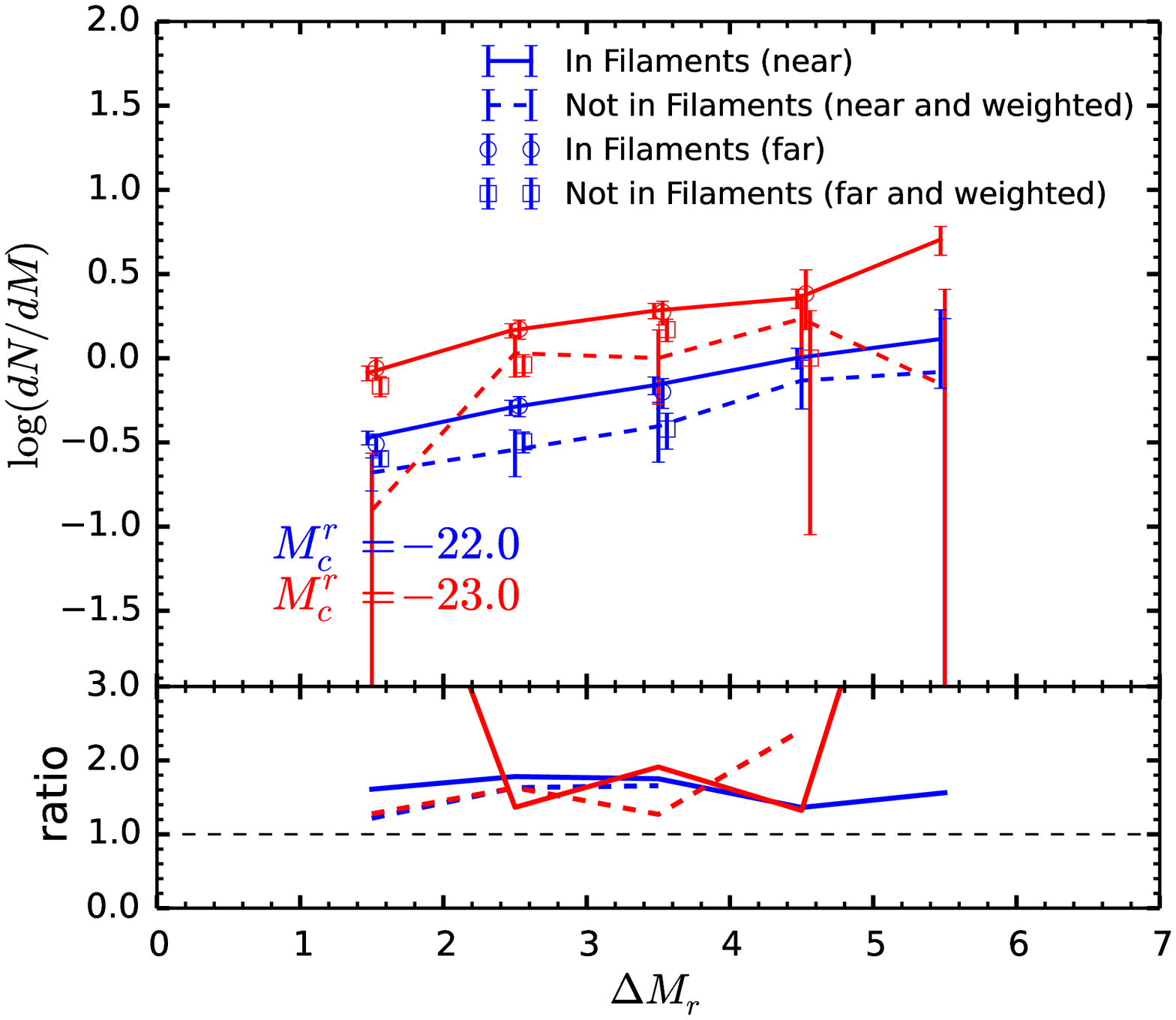}
      \caption{Similar to Figure~\ref{fig:fig3_lfw}, but where the weighting is
      done with respect to ``near'' and ``far'' subsamples. {\it Top:} We show
      the mean satellite LF for in-filaments galaxies in the near subsample
      (solid lines), in-filaments primaries in the far subsample (open circles),
      not-in-filaments primaries in near subsample (dashed lines) and
      not-in-filaments primaries in far subsample (open squares). The ratio of
      the two near (far) subsamples are shown in the lower panel as the solid
      (dashed) lines. {\it Bottom:} the satellite LF of the not-in-filament
      primaries is weighted as in Figure~\ref{fig:fig3_lfw}. The codes of
    colours are the same as in Fig.~\ref{fig:fig2_lf}.} \label{fig:lf_redshift}
  \end{figure}

  To further explore how the changing fractions of primaries in-filaments can
  influence the differences between the satellite LFs, we split the in-filaments
  and not-in-filaments primaries into ``near'' ($0.04<z<0.09$, $0.05<z<0.11$ for
  $\satmc=-22.0, -23.0$ respectively) and ``far'' ($0.09<z<0.14$, $0.11<z<0.15$)
  subsamples.  The resulting satellite LFs for these subsamples are shown in
  Fig.~\ref{fig:lf_redshift}. The satellites LFs for in filament and
  not-in-filament primaries are still significantly different for both near and
  far subsamples, although the differences in far subsamples are smaller than
  those in near subsamples.  This could be caused by the fact that the
  contamination in not-in-filaments primaries in far subsamples is higher than
  that in near subsamples, i.e. more in-filaments galaxies in far subsamples may
  be mis-classified as not-in-filaments. The weighted satellite LFs of galaxies
  not in filaments are shown in the two bottom panels of
  Fig.~\ref{fig:lf_redshift} as an attempt to minimise the redshift-bias between
  the in-filament galaxies and the not-in-filaments ones in both near and far
  subsamples. Due to the small number of primaries in the subsamples which are
  able to contribute the estimation of the weighted satellite LFs, the results
  are quite noisy. However the difference between the mean of satellite LFs in
  filaments and not in filaments can still be seen.

  \begin{figure}
    \plotone{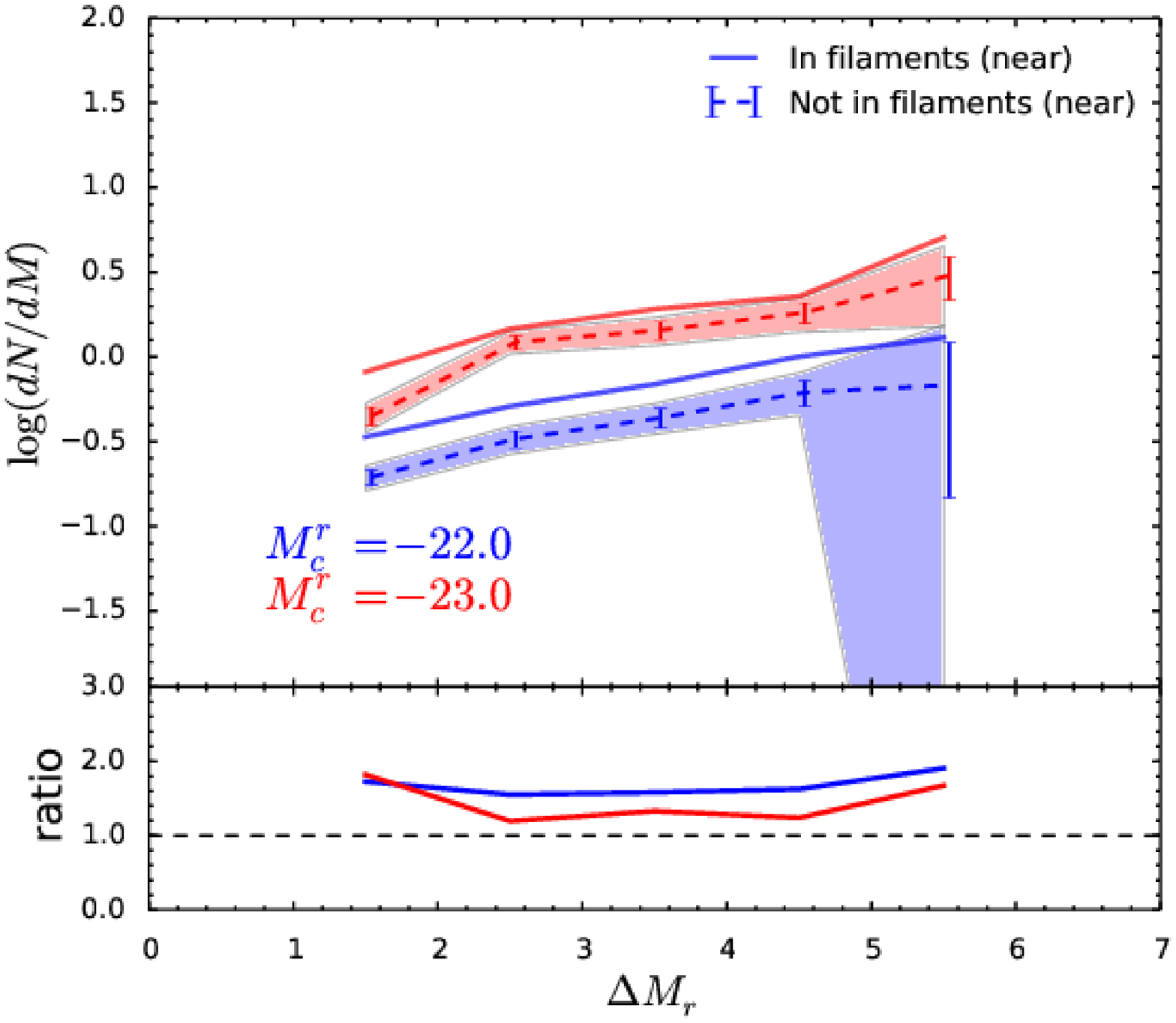}
    \plotone{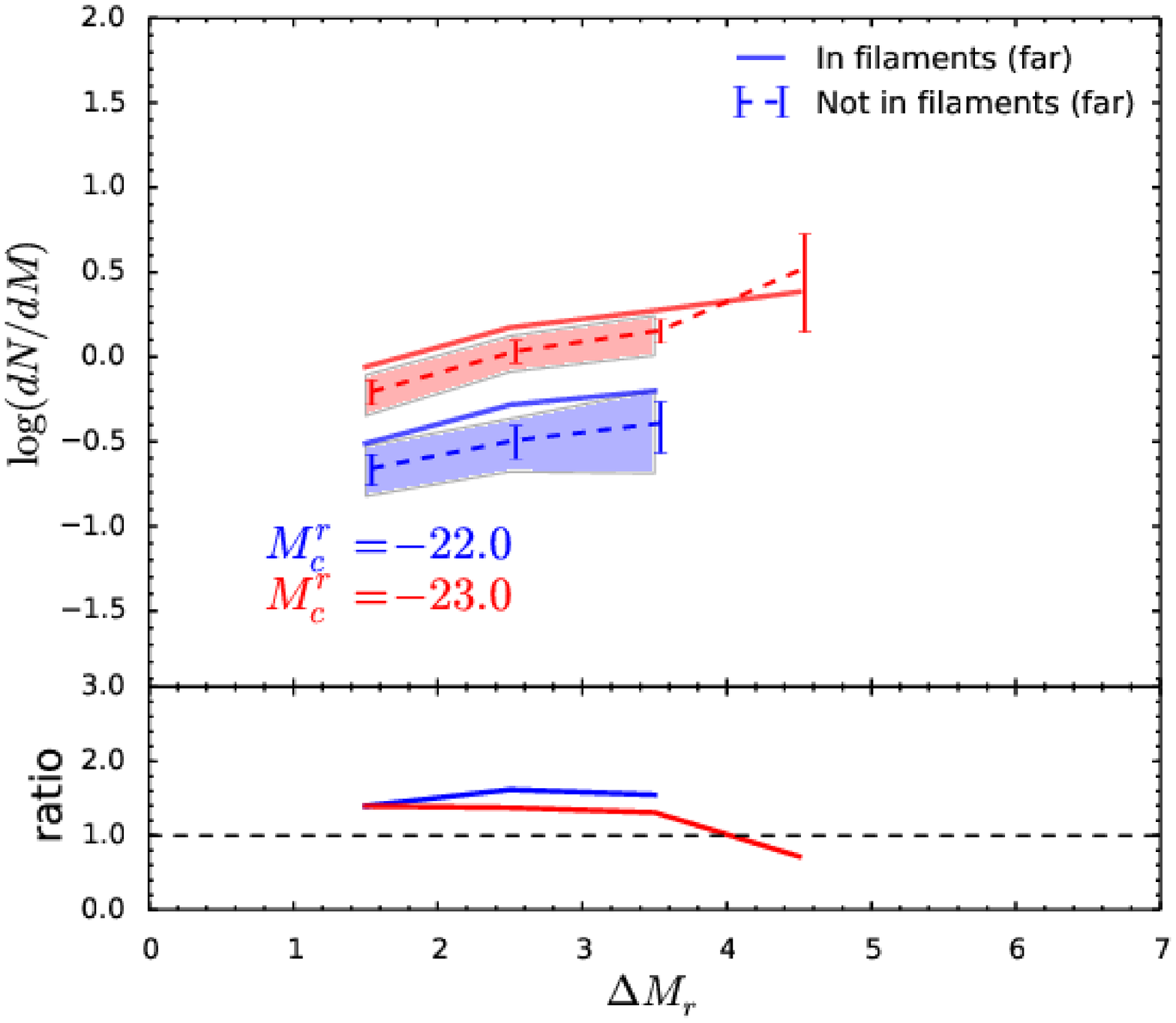}
    \caption{The mean satellite LFs composed by randomly selecting 500
      primaries from filament primaries (solid line) and not-in-filament
      primaries (dashed) in near (top) and far (bottom) samples. The error bars
      show the 1$\sigma$ standard deviation of the distribution of the mean
      satellite LFs. The shaded regions show the corresponding $5^{th}$ and
      $95^{th}$ quantiles.The small lower panels show the ratios of
      corresponding mean satellite LFs as in previous Figures. Curves are color
      coded by magnitude as in previous figures} \label{fig:lf_frac}
  \end{figure}

  In each magnitude bin, only around 15\% of isolated galaxies are found in
  filaments. We wish to check if the difference in the satellite LF is due to
  the variance of this small fraction.  We do so by comparing the mean
  in-filament satellite LFs to the mean of (a random set of) 500 not-in-filament
  satellite LFs. This set of 500 random satellite LFs matches both the redshift
  range and the number of the in-filaments galaxies. Again, we split these into
  near and far samples. Figure~\ref{fig:lf_frac} shows the results of comparing
  these. The difference between the mean satellite LFs of in-filaments galaxies
  and of a randomly drawn, equally sized and redshift distributed
  not-in-filaments sample  in both near and far ranges, is robust and
  statistically significant (at the $\sim 2\sigma$ level).

\begin{figure}
      \plotone{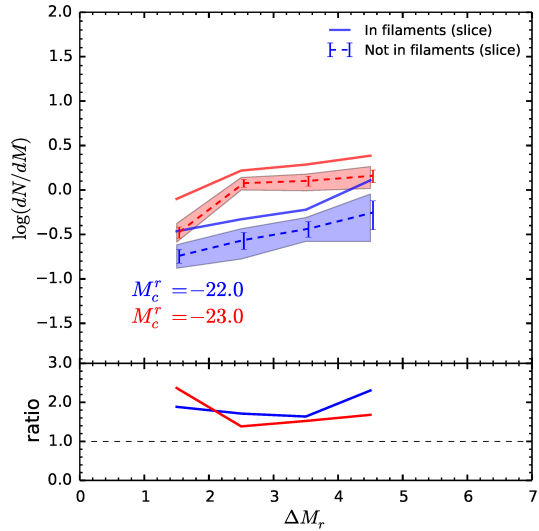}
      \caption{Same as Figure~\ref{fig:lf_frac}, but for a narrow redshift
      slices ($0.07<z<0.09$, $0.08<z<0.10$ for $M_{r}=-22.0, -23.0$
      respectively).}
      \label{fig:lf_frac_slice}
  \end{figure}

  Is our result valid if the redshift range is confined to small interval? In
  Figure~\ref{fig:lf_frac_slice} we show  the satellite LF in two relatively
  narrow redshift  slices ($0.07<z<0.09$, $0.08<z<0.10$ for $\satmc=-22.0,
  -23.0$). In these narrow redshift ranges, the fractions of in-filaments
  primaries varies very little. The comparison of satellite LFs is thus less
  biased by counting statistics. Moreover, bright satellites in these redshift
  slices are quite far from the faint limits of the SDSS. The estimation of the
  mean satellite LFs is therefore less affected by possible incompleteness. The
  results from these subsamples again show that the mean satellite LFs of
  in-filaments galaxies and not-in-filaments galaxies are significantly different.

  To summarize: the difference between the satellite LFs of in-filament and
  not-in-filament primaries are seen in a wide range of subsamples. The fact
  that the fraction of galaxies in-filaments varies with the redshift
  dramatically is thus insufficient to explain the difference in the satellite
  LF.

  \subsection{Control sample}

  \label{sect:controlsample}
  
  Besides the different redshift and colour distributions of the two samples,
  the classification of primaries as in-filaments and not-in-filaments could
  introduce other potential biases which can individually or together result in
  the differences seen here. To take such an effect into account, we build
  control samples from the not-in-filament catalogue. We then compare the mean
  satellite LFs estimated from galaxies in-filaments with a corresponding
  control samples of galaxies not-in-filaments.  For each primary found in a
  filament, we select a unique counterpart from the corresponding
  not-in-filament control sample. The counterpart is selected to have the same
  $g-r$ colour, visible axis ratio $b/a$, redshift and magnitude.  Given a
  filament primary, we select a counterpart by finding the not-in-filament
  galaxy which minimizes the following cost :  

   \footnotesize
    \begin{equation}
        C=\sqrt{\left(\frac{\Delta z}{0.036}\right)^2 + \left(\frac{\Delta
        M_{r}}{0.75}\right)^2 + \left(\frac{\Delta(g-r)}{0.15}\right)^2 + 
        \left(\frac{\Delta(b/a)}{0.25}\right)^2},
    \end{equation}
  \normalsize

  where the quantities $\Delta z$, $\Delta M_{r}$, $\Delta(g-r)$, and
  $\Delta(b/a)$ are the difference between the redshift, $r$ band magnitude, $g-r$
  colour and axis ratio between the filament primary and not-in-filament
  candidate.  The denominators of each term are the scaling factors that
  correspond to the standard deviation of distribution of each quantity.

\begin{figure}
    \plotone{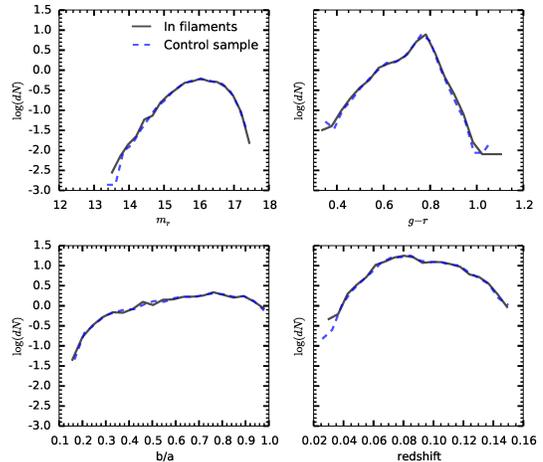}
    \caption{A comparison of the properties of filament primaries (solid
    lines) with their not-in-filaments twins that constitute our control
    sample (dashed lines) for magnitude bin $\satmc = -22.0$.} 
    \label{fig:comp_p}
\end{figure}

\begin{figure}
      \plotone{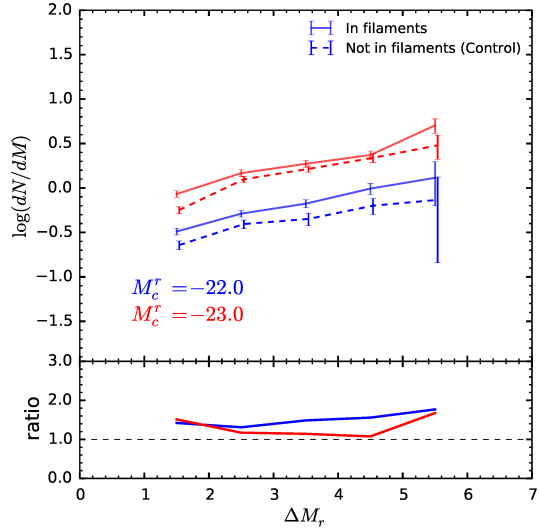}
        \caption{The mean satellite LFs of in-filaments primaries (solid lines)
        and a control sample of ``twins'' chosen from the not-in-filaments sample
      (dashed lines) according to Equation~2, color coded by magnitude. The ratio of
      the satellite LFs of in-filaments and not-in-filaments are shown in the small
      lower panel.}  
      \label{fig:lf_control}
  \end{figure}

  The general properties of our control sample drawn from primaries
  not-in-filaments is shown in Figure~\ref{fig:comp_p}. Note that our technique of
  minimizing the cost $C$ is successful: Figure~\ref{fig:comp_p} shows little
  difference between in filament primaries and the control sample. This makes the
  comparison of the mean satellite LFs from these two groups less biased.
  Figure~\ref{fig:lf_control} shows the mean satellite LFs of in-filaments and
  not-in-filaments galaxies of the control samples, which again confirms the
  differences between the satellite LFs of in-filaments and not-in-filaments
  galaxies is not caused by a possible bias in the distribution of magnitudes,
  redshift, colour, or visible axis ratio of the primary galaxies.

\section{Summary and Conclusions}
  We have examined the luminosity function (LF) of satellites close to isolated
  primary galaxies in the SDSS. Isolated primary galaxies have been split by
  filamentary environment into two camps: those primaries in and not in filaments.
  Background galaxies are subtracted statistically and the mean satellite LF is
  computed by stacking all centrals of a given absolute $r$ band magnitude.

  Our results indicate that primaries in filaments have more satellites than
  those that are not found in filamentary environments. This is most evident for
  the brightest satellites but is true for satellites up to 4 magnitudes fainter
  of their host. Except for the faintest magnitude bin (where the
  signal-to-noise is too low to judge), in-filament primaries have a factor of
  $\sim 1.5$--$2$ more bright satellites than primaries not in filaments. This
  slightly varies with redshift possibly because the contamination in the sample
  of galaxies not in filaments increases with the redshift.

  We have examined if the difference in the satellite LF is due to other
  controlling factors including differences in colour, density, redshift
  distribution or variance of small sample sizes. None of these control factors
  can explain the result found here. The differences exhibited are statistically
  significant and robust and thus reflect an inherent difference in the satellite
  population of galaxies in and not in filaments. 

 We have also examined the mean radial distribution of satellites brighter than
 a $r$-band magnitude of $-20$ in Appendix~\ref{appendix:profile} and found no
 significant difference in the projected spatial distribution of satellites in
 or not in filaments in spite of the obvious difference between the total
 numbers of satellites.

  Previous numerical studies have shown that the abundance of dark matter
  subhaloes may depend on the environment: haloes in filaments may have more
  subhaloes than those in other cosmic web environments \citep[Cautun et al. in
  preparation;][]{ney14}.  The properties of subhaloes in filaments can be also
  different from those in the other environments and this may affect galaxy
  formation since the satellite LF is a result of the gas physics that regulate
  star formation in small haloes. Our results thus suggest that galaxy formation
  itself may be more efficient in subhaloes that are born in and accreted
  through filaments. The filamentary environment may be a crucial component of
  galaxy formation on such small scales.

\acknowledgments
ET acknowledge the ESF grants MJD272, IUT40-2 and the European Regional
Development Fund. NIL acknowledges a grant from the \textit{Deutsche
Forschungs Gemeinschaft}.
Funding for SDSS-III has been provided by the Alfred P. Sloan Foundation, the
Participating Institutions, the National Science Foundation, and the U.S.
Department of Energy Office of Science. The SDSS-III web site is
http://www.sdss3.org/. SDSS-III is managed by the Astrophysical Research
Consortium for the Participating Institutions of the SDSS-III Collaboration
including the University of Arizona, the Brazilian Participation Group,
Brookhaven National Laboratory, Carnegie Mellon University, University of
Florida, the French Participation Group, the German Participation Group, Harvard
University, the Instituto de Astrofisica de Canarias, the Michigan State/Notre
Dame/JINA Participation Group, Johns Hopkins University, Lawrence Berkeley
National Laboratory, Max Planck Institute for Astrophysics, Max Planck Institute
for Extraterrestrial Physics, New Mexico State University, New York University,
Ohio State University, Pennsylvania State University, University of Portsmouth,
Princeton University, the Spanish Participation Group, University of Tokyo,
University of Utah, Vanderbilt University, University of Virginia, University of
Washington, and Yale University.

\appendix

\begin{figure*}
  \begin{center}
    \includegraphics[width=75mm]{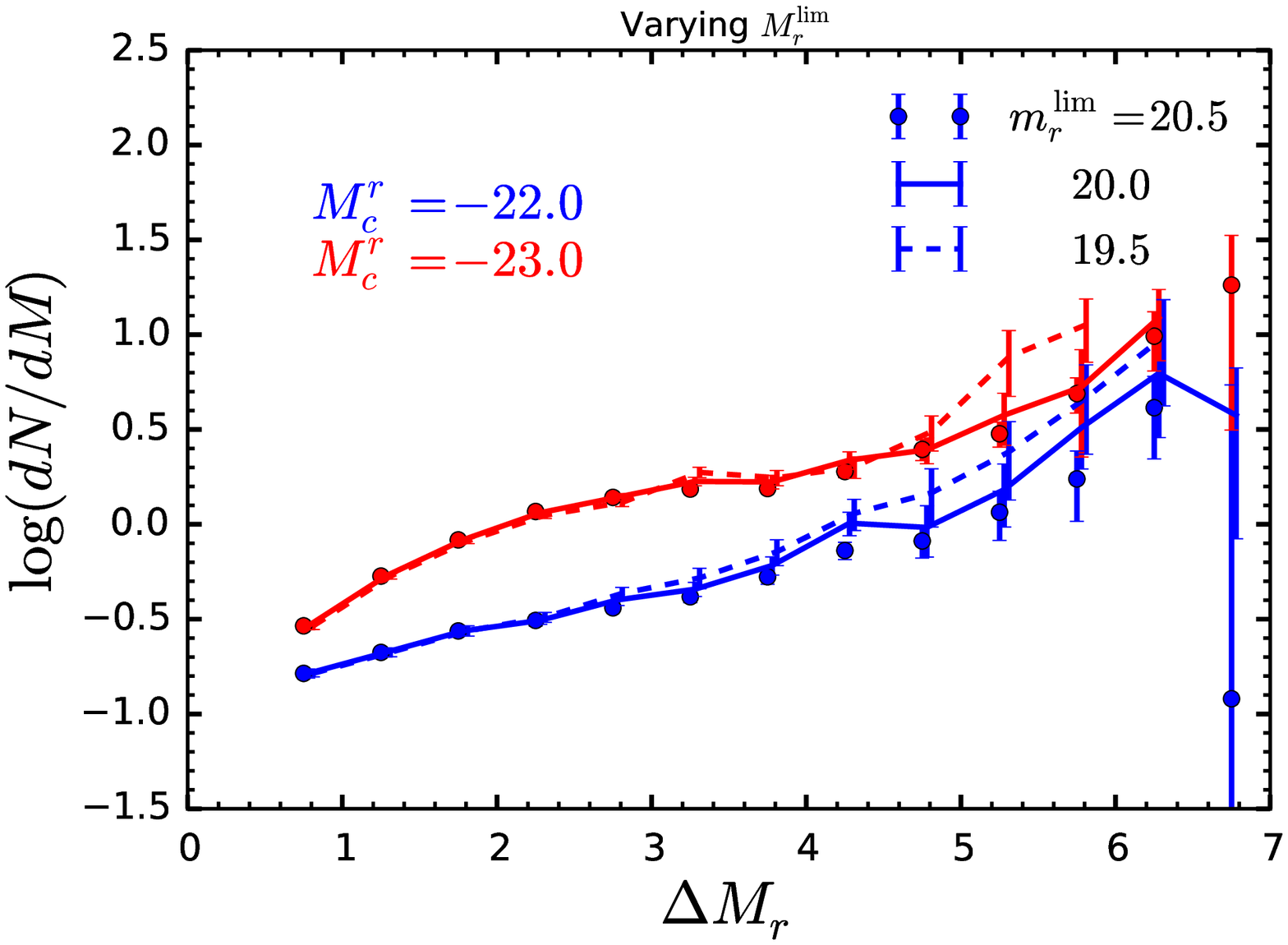}
    \includegraphics[width=75mm]{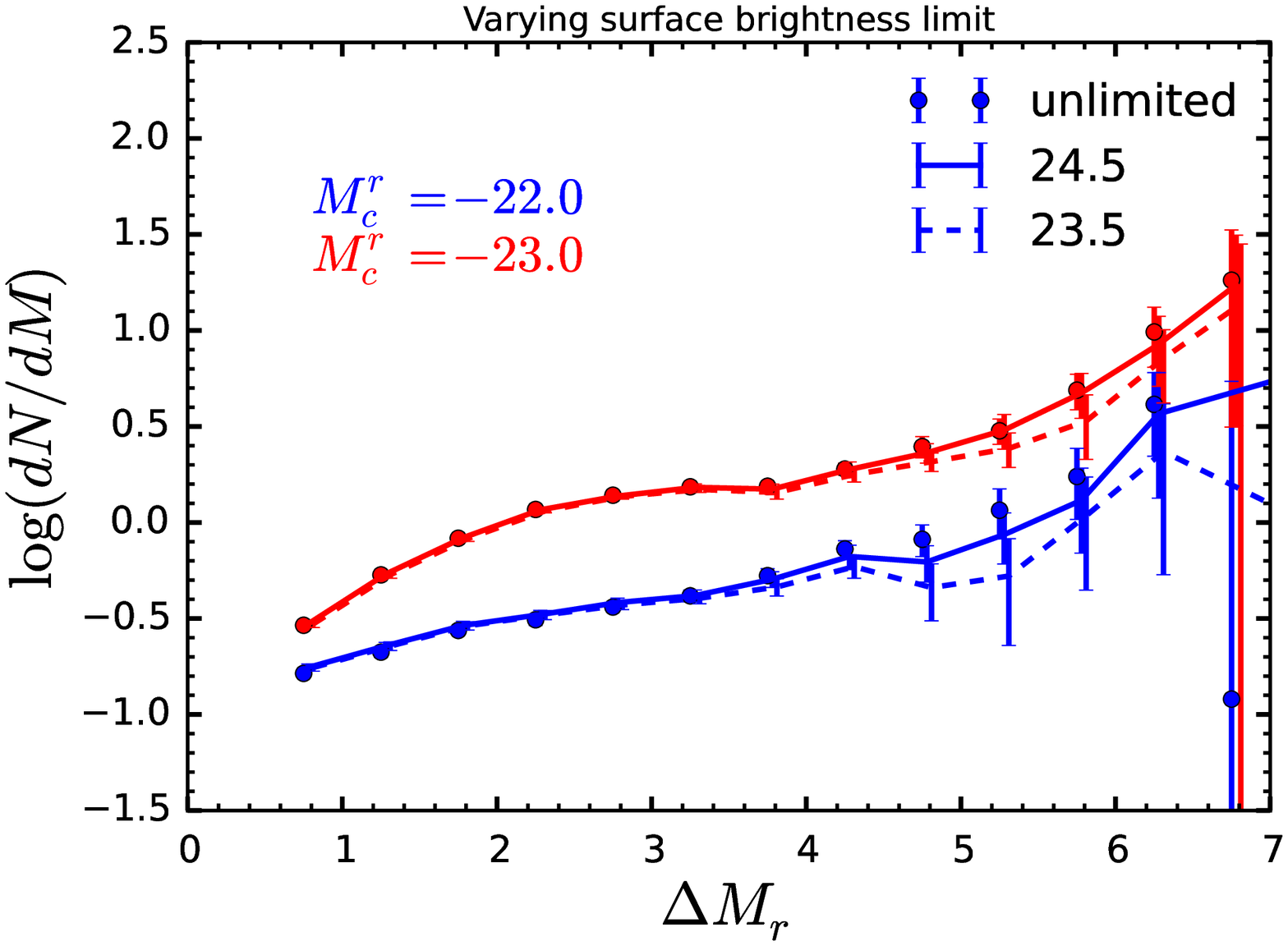}
    \includegraphics[width=75mm]{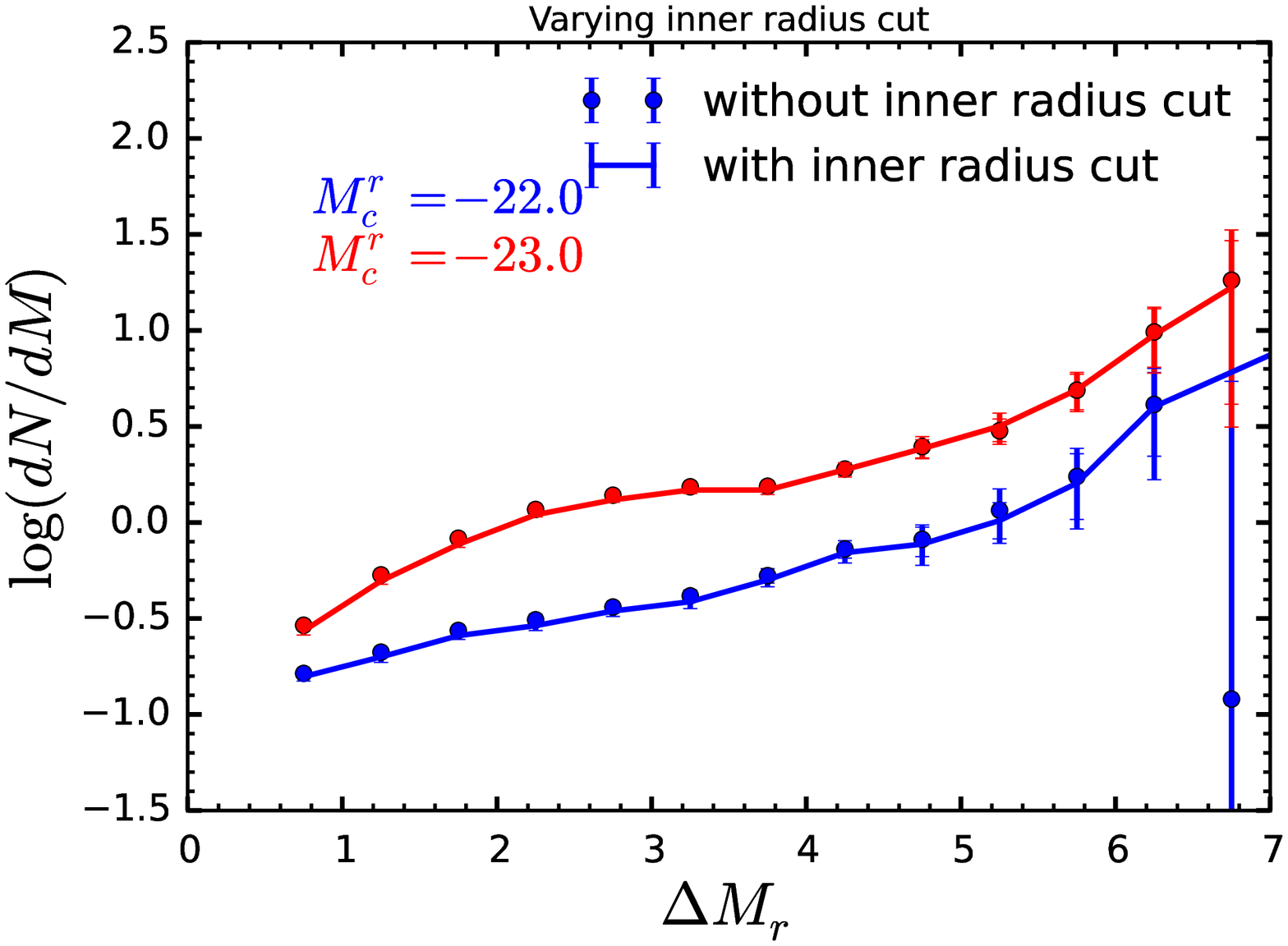}
    \includegraphics[width=75mm]{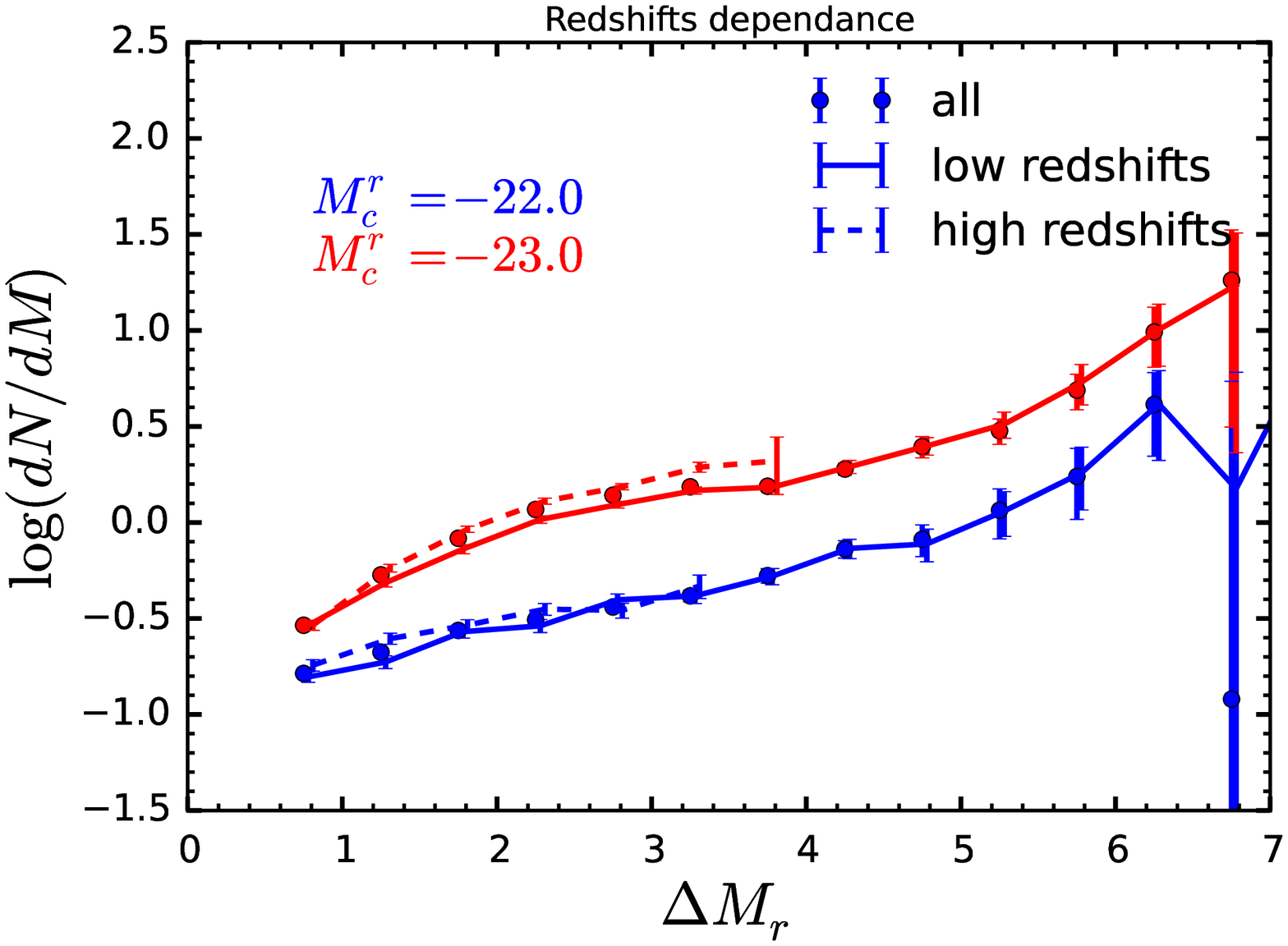}
  \end{center}
    \caption{The effect on the satellite LFs of varying the magnitude limit,
    $m^{r}_{\rm lim}$ (upper left), the surface brightness limit (upper right),
    the inner radius (bottom left), and splitting the primary galaxies in the
    subsamples at low and high redshift (bottom right). The satellite LF with
    different parameters and default parameter are shown as solid, dashed lines
    and filled points respectively. The color coding is same as in previous
    figures.}
    \label{fig:test}
\end{figure*}

\section{Test of the estimation of satellite LFs}
\label{appendix:test}

In Figure~5 of \cite{my11}, the estimated satellite LFs were tested and found to
be robust to changes in the values of selection parameters: $\satdmbin$,
$\satdmf$, $\satdzs$, $\satap$ $m_{\rm v}^{\rm lim}$. Here we perform a few more
tests to see if our estimated satellite LFs are also robust in terms of other
possible incompleteness or biases. In the upper right panel of Fig.~\ref{fig:test} (titled:
``varying the magnitude limit $m_{r}^{\rm lim}$''), we reduce the faint
magnitude cut of the input catalogue used for searching for satellites. The
resulting satellite LFs at the bright end is nearly identical. Such a cut causes
only small variations in the faint end of the satellite LFs. A similar situation
is seen if we vary the surface brightness cut for the input catalogue (upper
right of Fig.~\ref{fig:test}).  Figure~5 of \cite{my11} shows that the galaxy catalogue is
not complete due to preferentially missing low surface brightness galaxies. With
the bright magnitude cut, such incompleteness is greatly reduced. Therefore
varying the surface brightness cut does not affect the satellite LFs.  In the
region close to the primary galaxies, spiral arms fragments could be
occasionally erroneously misclassified as separate galaxies. Moreover the
completeness in the region close to primary is poor since the light of the
central galaxy dominates this region. For such incompleteness, we compare the
satellite LFs estimated with and without excluding the galaxies within a radius
of 1.5 times the Petrosian $R_{90}$ of the primary galaxies, shown in the lower
left panel in Fig.~\ref{fig:test} (entitled ``varying the inner radius cut''). The result
shows that such incompleteness is not important in the computation of the
satellite LF. This is because the $R_{90}$ is small. We thus infer that objects
found at these distances contribute little to the mean satellite LFs. In the
bottom right panel of Fig.~\ref{fig:test}, we test the dependence of the estimated satellite
LFs on the redshift of primary galaxies. The primary galaxies are split into
subsamples at low and high redshift according to the median of the redshift
distribution of all isolated galaxies in the magnitude bins. The satellites LFs
estimated from the sample of primary galaxies at low redshift are compared with
those from the sample of primary galaxies at high redshift, shows that only for
the $\satmc=-23.0$ bin is the satellite LFs slightly different.  However, such
variation is not enough to account for the difference between the satellite LFs
of in-filaments galaxies and not-in-filaments galaxies. For tests of the
dependence on the redshift of primary galaxies we refer the reader to the
Section~\ref{sec:redshift} for more details.

In summary, our tests show that our estimated satellite LFs of isolated
primaries, especially the bright end, are quite robust to parameter selections,
incompleteness or biases. There is thus no evidence that the differences in the
satellite LFs of in-filaments and not-in-filament galaxies is anything other
than a real physical effect.

\section{Projected radial distribution}
  \label{appendix:profile}
\begin{figure}
    \plotone{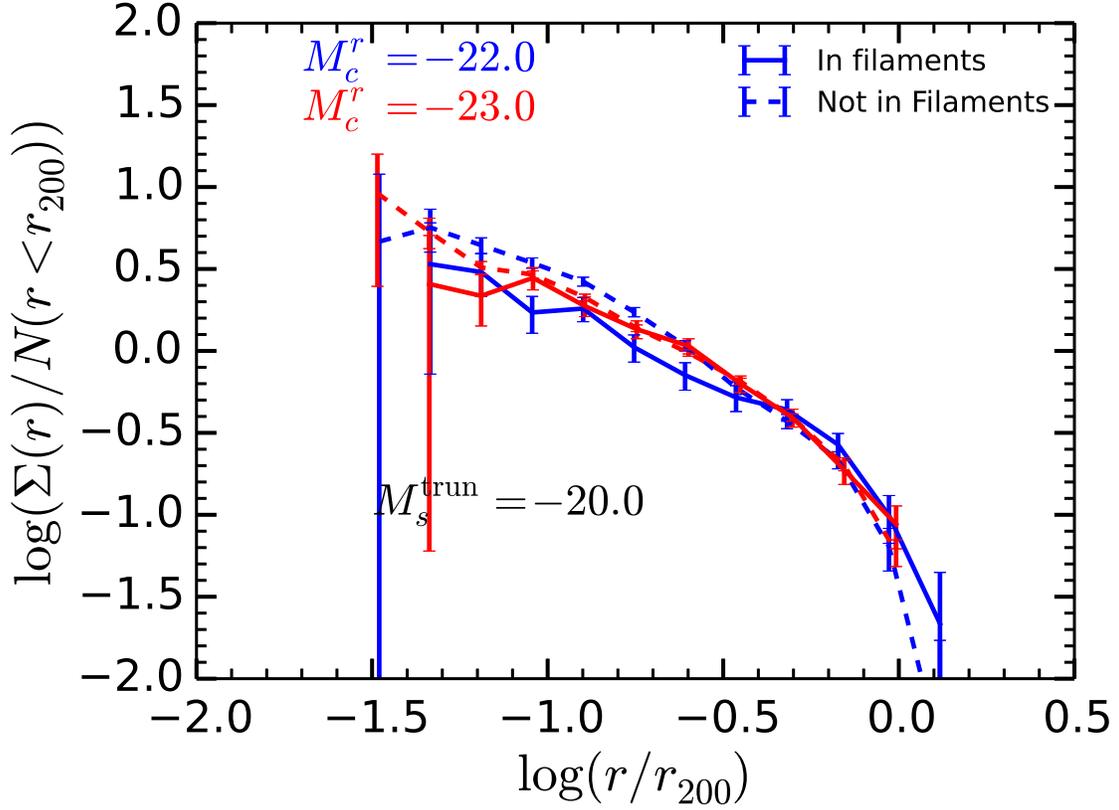}
    \caption{The scaled projected density profiles of satellites brighter than
    $-20$ magnitude in $r$ band for the primary galaxies in filaments and
    not in filaments. The codes of legends are the same as in Fig.~\ref{fig:fig2_lf}.}
    \label{fig:fig4_pro}
\end{figure}

  The difference at the bright end of the satellite LFs for these two subsamples
  is most significant in the brightest two magnitude bins. We therefore show the
  radial distribution of satellites in and not in filaments (for $\satmc =
  -22.0, -23.0$) in Figure~\ref{fig:fig4_pro}.  In spite of the fact that the
  total number of satellites depends strongly on whether the primary is inside a
  filament or not, the mean radial distributions of satellites in these two
  environments are remarkably similar to each other. However for the magnitude bin
  $\satmc = -22.0$, the radial distribution of satellites brighter than $-20$ magnitude
  in $r$ band seems slightly less centrally concentrated  in filaments than
  not-in-filaments. This may be caused by the limited volume of the primary
  galaxies in filaments, since the result is relative noisy. 

\end{document}